\documentclass[11pt,showfigures]{natneurosci_sm}
\usepackage{geometry}                
\usepackage{graphicx}
\usepackage{amssymb}
\usepackage{amsmath}
\usepackage{color}

\usepackage{authblk}
\usepackage[superscript,nomove]{cite}

\usepackage{setspace}

\newcommand{\mc}{\mathcal}

\newcommand{\transpose}{\mathsf{T}}

\title{Stimulation-based control of dynamic brain networks}
\author[1,2,3]{Sarah Feldt Muldoon}
\author[4]{Fabio Pasqualetti}
\author[1,5]{Shi Gu}
\author[6]{Matthew Cieslak}
\author[6]{Scott T. Grafton}
\author[2,6,1]{Jean M. Vettel}
\author[1,7]{Danielle S. Bassett}
\affil[1]{Department of Bioengineering, University of Pennsylvania, Philadelphia, PA 19104}
\affil[2]{US Army Research Laboratory, Aberdeen Proving Ground, MD 21005}
\affil[3]{Department of Mathematics and Computational and Data-Enabled Science and Engineering Program, University at Buffalo, SUNY, Buffalo, NY 14260}
\affil[4]{Department of Mechanical Engineering, University of California, Riverside, CA 92521}
\affil[5]{Applied Mathematics and Computational Science Graduate Program, University of Pennsylvania, Philadelphia, PA 19104}
\affil[6]{Department of Psychological and Brain Sciences, University of California, Santa Barbara, Santa Barbara, CA 93106}
\affil[7]{Department of Electrical and Systems Engineering, University of Pennsylvania, Philadelphia, PA 19104}

\begin{document}

\maketitle

\begin{abstract}
The ability to modulate brain states using targeted stimulation is increasingly being employed to treat neurological disorders and to enhance human performance.  Despite the growing interest in brain stimulation as a form of neuromodulation, much remains unknown about the network-level impact of these focal perturbations.  To study the system wide impact of regional stimulation, we employ a data-driven computational model of nonlinear brain dynamics to systematically explore the effects of targeted stimulation.  Validating predictions from network control theory, we uncover the relationship between regional controllability and the focal versus global impact of stimulation, and we relate these findings to differences in the underlying network architecture.  Finally, by mapping brain regions to cognitive systems, we observe that the default mode system imparts large global change despite being highly constrained by structural connectivity. This work forms an important step towards the development of personalized stimulation protocols for medical treatment or performance enhancement.
\end{abstract}

\newpage
\clearpage

\begin{introduction}

Brain stimulation is increasingly used to diagnose \cite{Volz:2015de}, monitor \cite{Helfrich:2012ho}, and treat neurological \cite{Tierney:2013wt} and psychiatric \cite{Temel:2012gr} disorders. Non-invasive stimulation, such as transcranial magnetic stimulation (TMS) or transcranial direct current stimulation (tDCS), is used, for example, in epilepsy \cite{Terra:2013cm,ChristopherMDeGiorgio:2013bi}, stroke \cite{DiLazzaro:2013ga}, attention deficit hyperactivity disorder \cite{Helfrich:2012ho}, tinnitus \cite{Vanneste:2013if}, headache \cite{Jurgens:2013io}, aphasia \cite{Shah:2013dy}, traumatic brain injury \cite{Bonni:2013jb}, schizophrenia \cite{Hasan:2013uy}, Huntington's disease \cite{Berardelli:2013gb}, and pain \cite{Andrade:2013bz}, while invasive deep brain stimulation (DBS) is approved for essential tremor and Parkinson's disease and is being tested in multiple Phase III clinical trials in major depressive disorder, Tourette's syndrome, dystonia, epilepsy, and obsessive-compulsive disorder \cite{Lozano:2013ej}. In addition to its clinical utility, emerging evidence suggests that stimulation can also be used to optimize human performance in healthy individuals \cite{Meinzer:2014cf,Ferreri:2013kn}, potentially by altering cortical plasticity \cite{Ferreri:2013kn}.

Despite its broad utility, many engineering challenges remain \cite{Johnson:2013gf}, from the optimization of stimulation parameters to the identification of target areas that maximize clinical utility \cite{Chaieb:2009dk}. Critically, an understanding of the local effects of stimulation on neurophysiological processes -- and the downstream effects of stimulation on distributed cortical and subcortical networks -- remains elusive \cite{Hess:2013de}. This gap has motivated the combination of stimulation techniques with various recording devices (PET \cite{Grafton:2006jk}, MEG \cite{Soekadar:2013jo}, fast optical imaging \cite{Parks:2012kh}, EEG \cite{Ferreri:2013kn}, and fMRI \cite{PascualLeone:2011fk,Leitao:2013fv}) to monitor the effects of stimulation on cortical activity \cite{LevitBinnun:2007iq,Grefkes:2011hn,Arzouan:2014bf}. In this context, it has become apparent that there is a critical need for biologically informed computational models and theory for predicting the impact of focal neurostimulation on distributed brain networks, thereby enabling the generalization of these effects across clinical cohorts as well as the optimization of stimulation protocols \cite{Bestmann:2013bd}, including refinements for individualized treatment and personalized medicine.

To meet this fast-growing need, we utilize network control theory \cite{Gu:2015dr} to understand and predict the effects of stimulation on brain networks. The advent of network theory as a ubiquitous paradigm to study complex engineering systems as well as social and biological models has changed the face of the field of automatic control and redefined its classic application domain. \emph{Control} of a network refers to the possibility of manipulating local interactions of dynamic components to steer the global system along a chosen trajectory. Network control theory offers a mechanistic explanation for how specific regions within well-studied cognitive systems may enable task-relevant neural computations: for example, activity in the primary visual cortex can initiate a trajectory of processing along the extended visual pathway (V2, V3, etc.), driving visual perception and scene understanding. More broadly, network control theory also offers a mechanistic framework in which to understand clinical interventions such as brain stimulation, which elicits a strategic functional effect within the network for synchronized neural processing. In both cases (regional activity and stimulation), it is critically important to understand how underlying structural connectivity can constrain or modulate the functional effect of regional alterations in activity. Yet, the application of control-theoretic techniques to brain networks is underexplored, even though the questions posed by the neuroscientific community are uniquely suited to the application of these tools \cite{Gu:2015dr}.

In a novel approach to link network control theory and brain dynamics, we examine the effects of regional stimulation on brain states using a nonlinear meso-scale computational model, built on data-driven structural brain networks.  We demonstrate that the dynamics of our model is highly variable across subjects, but highly reproducible across multiple scans of the same subject. We confirm the pragmatic utility of network control theory for nonlinear systems, extending previous work on linear approaches \cite{Gu:2015dr}, and show that general control diagnostics (average and modal controllability) are strongly correlated with the density of structural connections linking brain regions. Finally, we investigate the interplay between functional and structural effects of stimulation by examining how the global functional activity across brain regions is modulated by region-specific stimulation (a region's functional effect) and whether the region's structural connectivity accounts for its influence on the larger brain network (structural effect).  Results show that the default mode system and subcortical regions produce the strongest functional effects; the subcortical structures display weak structural effects, being diversely connected across many cognitive systems.  Collectively, our results indicate the value of data-driven, biologically motivated brain models to understand how individual variability in brain networks influences the functional effects of region-specific stimulation for clinical intervention or cognitive enhancement.

\end{introduction}

\begin{results}

To systematically assess the effects of brain stimulation and its utility in control, we begin by building a spatially embedded nonlinear model of brain network dynamics. Structural brain networks are derived from diffusion spectrum imaging (DSI) data acquired in triplicate from $8$ healthy adult subjects. We perform diffusion tractography to estimate the number of streamlines linking $N=83$ large-scale cortical and subcortical regions extracted from the Lausanne atlas \cite{Hagmann:2008gd} and summarize these estimates in a weighted adjacency matrix whose entries reflect the density of streamlines connecting different regions. Finally, we model regional brain activity using biologically motivated nonlinear Wilson-Cowan oscillators \cite{Wilson:1972kj}, coupled through these data-derived structural brain networks. We quantify brain states based on functional connectivity obtained from pairwise correlations between simulated regional brain dynamics (Fig.~\ref{fig1} and Methods). 

\begin{figure}
\centerline{\includegraphics{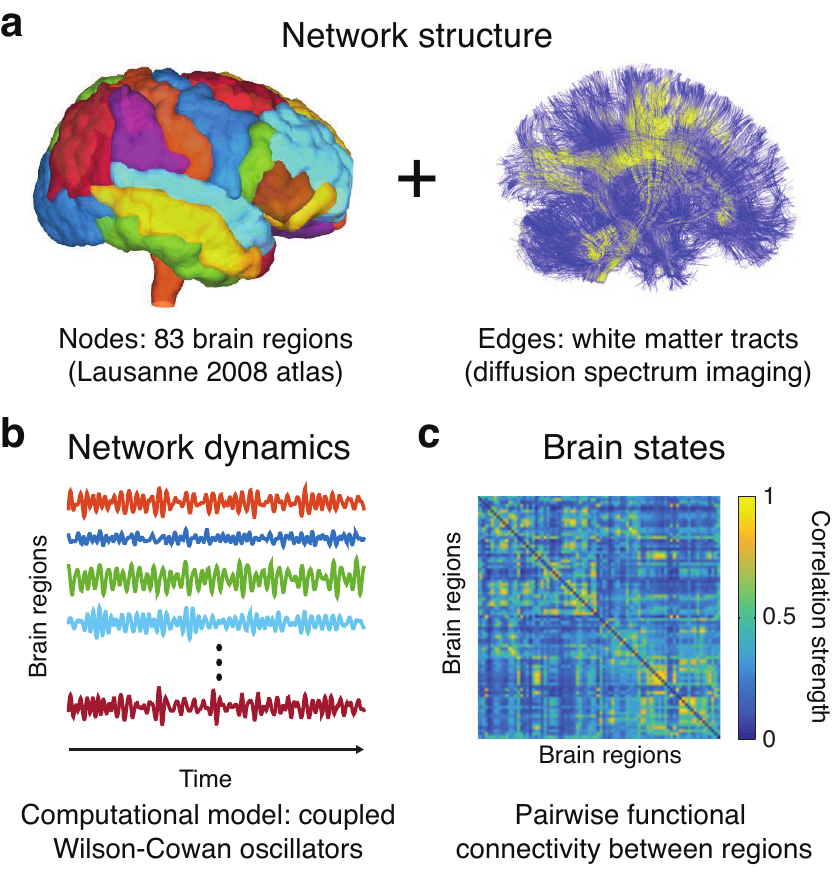}}
\caption{\textbf{Building nonlinear brain networks.}  (\textbf{a}) Subject-specific structural brain networks are built based on a parcellation of the brain into 83 anatomically defined brain regions (network nodes) with connections between regions given by the density of streamlines linking them. (\textbf{b}) The dynamics of each region are represented by a single Wilson-Cowan oscillator, and these oscillators are coupled according to the structural connectivity of a single subject.  (\textbf{c}) Brain states are quantified by calculating the pairwise functional connectivity between brain regions.
\label{fig1}}
\end{figure}

\subsection*{Inter-subject variability of brain dynamics}
Our broad goal is to understand the role of regional stimulation to differentially control brain dynamics. Theoretical predictions of regional controllability can be developed based on the underlying structural connectivity of the computational model \cite{Gu:2015dr}. Given our subject-specific data-driven approach, it is therefore important to understand how variability between structural brain networks affects the dynamics of our model.  Wilson-Cowan oscillators are a biologically driven mathematical model of the mean-field dynamics of a spatially localized population of neurons \cite{Wilson:1972kj,Wilson:1973uv}, modeled through equations governing the firing rate of coupled excitatory ($E$) and inhibitory ($I$) neuronal populations (Methods).  Here, we measure a single brain region's dynamics by the firing rate of the excitatory population.   An important feature of these Wilson-Cowan oscillators is that an uncoupled oscillator can exhibit one of three states, depending upon the amount of external current applied to the system (Fig.~\ref{fig2}a-b).  When no external current is applied ($P=0$), the system relaxes to a low fixed point (Fig.~\ref{fig2}a-b).  For moderate amounts of applied current, the oscillator is pushed into an oscillatory limit cycle, and if sufficiently high amounts of current are applied, the system settles at a high fixed point.

For this system of coupled oscillators, brain regions (oscillators) can receive current from an external input (stimulation, i.e., $P > 0$) or from the activity of other brain regions to which they are connected.  The global coupling parameter, $c_5$, therefore serves to govern the global state of the system by regulating the overall amplitude of current transmitted between brain regions.  For low values of $c_5$, the system will fluctuate around the low fixed point, whereas for high values of $c_5$, the system will transition into the oscillatory (limit cycle) regime.  For a given structural connectivity, we can systematically increase the global coupling parameter and record the value at which the system transitions from the low fixed point to the oscillatory regime.  Individual differences in structural connectivity can cause this transition to occur at different points in the parameter space, and we therefore use this point of transition to assess the sensitivity of our model to inter subject differences in the structural connectivity.
	
By measuring the point of oscillatory transition using structural connectivity matrices obtained from each of three scans for eight subjects, we see in Fig.~\ref{fig2}c that model dynamics are highly reproducible within scans of a single subject, but show variability across subjects.  We quantify the within versus between subject reproducibility using the intraclass correlation coefficient ($ICC$).  We see high reproducibility within scans of a single subject (Fig.~\ref{fig2}d, $ICC=0.826$) and low reproducibility between different subjects ($ICC=-0.006$), which also corresponds to a low within subject variance ($V=0.0019$) and high between subject variance ($V=0.0143$).  Due to the high within subject reproducibility across the three scans of each subject, the remaining findings are presented at the subject level by averaging the results over simulations derived from each of the three single subject scans.

\begin{figure}
\centerline{\includegraphics{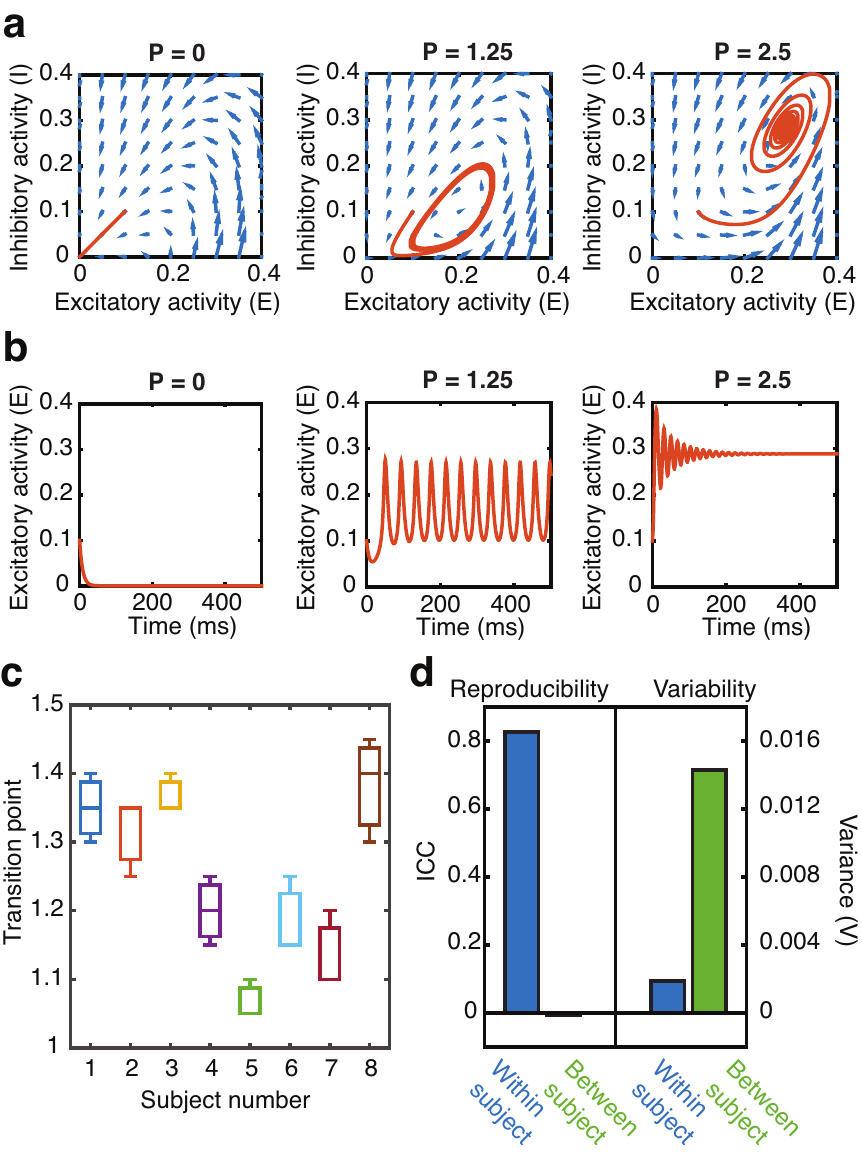}}
\caption{\textbf{Nonlinear brain dynamics and variability.}  (\textbf{a}) Excitatory-inhibitory phase space plots depicting behavior for a single Wilson-Cowan oscillator in the presence of no external current input (left; $P=0$; low-fixed point), moderate external current input (middle; $P=1.25$; limit cycle), and high external current input (right; $P=2.5$; high fixed point).  All simulations are started with initial conditions $E=0.1$, $I=0.1$. (\textbf{b}) The corresponding firing rate of the excitatory population plotted as a function of time for the simulations depicted in \emph{(a)}.  (\textbf{c}) Box plots showing the value of global coupling parameter at which the system transitions from the low fixed-point state to the oscillatory regime for models derived from three different structural scans from each of eight subjects.  (\textbf{d})  Within and between subject reproducibility (left) and variability (right) for the data shown in \emph{(c)}.  Reproducibility is measured by the intraclass correlation coefficient and is high within subjects, but low between subjects.  This is additionally reflected in the low within subject variability, measured as the average variance, and high between subject variability.
\label{fig2}}
\end{figure}

\subsection*{Regional controllability predicts the functional effect of stimulation}
In order to elucidate the role of regional stimulation to differentially control brain dynamics, we first turn to predictions made using linear network control theory \cite{Gu:2015dr}.  Linear network control theory assumes a simplified linear model of network dynamics and computes controllability measures based upon the topological features of the structural network architecture. Here, we assess two different types of regional controllability derived from our structural brain networks: average controllability and modal controllability.  Regions with high average controllability are capable of moving the system into many easy to reach states with a low energy input, whereas regions with high modal controllability can move the system into difficult to reach states but require a high energy input (see Methods for detailed descriptions of controllability measures and the Supplementary Information and Fig.~S1 for their relationship to the steady state network response from regional stimulation with a constant current input).  As previously described \cite{Gu:2015dr}, we observe a strong correlation between regional degree and average controllability and a strong inverse correlation between regional degree and modal controllability that is robust across structural networks derived from all subjects (Fig.~\ref{fig3}).

\begin{figure}
\centerline{\includegraphics{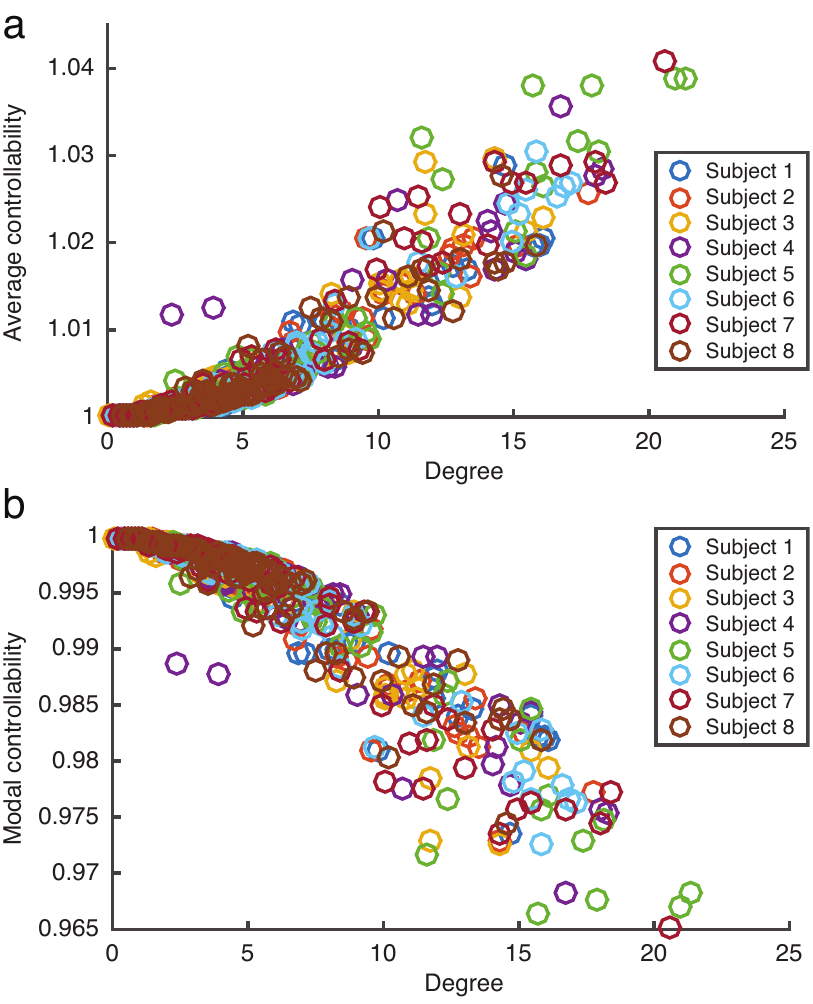}}
\caption{\textbf{Linear regional controllability.}  (\textbf{a}) Average controllability plotted as a function of regional degree for each of the 83 brain regions.   (\textbf{b}) Average controllability plotted as a function of regional degree.  Controllability predictions were performed for each of the three scans for each subject and the data points reflect controllability and degree values averaged across scans.
\label{fig3}}
\end{figure}

We predict that brain regions with a high average controllability have the ability to impart large changes in network dynamics, easily moving the system into many nearby states.  However, these predictions are made under the assumption of linear dynamics, and we know that the brain is in fact a highly nonlinear system.  Our modeling approach allows us to directly test the validity of these linear controllability predictions in a nonlinear setting by systematically studying the effects of focal stimulation to brain regions and studying how the system moves.

Using our computational model, we select the value of global coupling that places our brain model just before the transition to the oscillatory regime such that all brain regions are fluctuating near the low fixed point.  We then select a single brain region and add an external stimulating current that brings the selected region into its oscillatory state (Fig.~\ref{fig4}a).  We can quantify the changes in brain state due to this stimulation by computing the functional matrix of pairwise correlations between brain region dynamics during a period before stimulation occurs. We compare this pre-stimulation matrix to the functional matrix obtained during the stimulation period. By taking the difference between the functional brain state before and during regional stimulation, we measure the distance that the system moves.  As expected, stimulation of low controllability regions produces smaller changes in the functional brain state than stimulation of high controllability regions (Fig.~\ref{fig4}b-c).

\begin{figure}
\centerline{\includegraphics{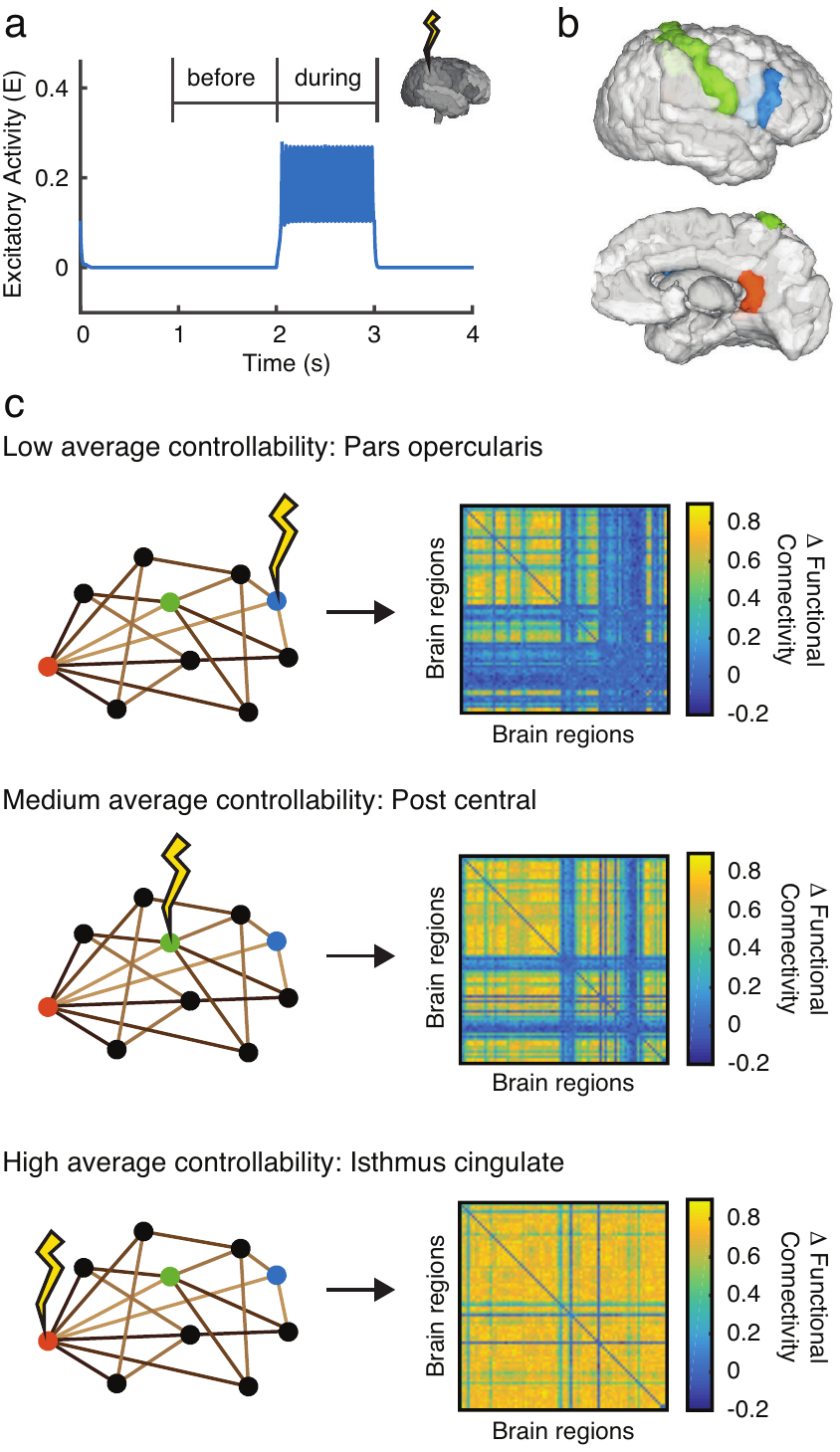}}
\caption{\textbf{Regional stimulation.}  (\textbf{a}) Stimulation of a single region pushes the region from fluctuations around its low fixed point to the oscillatory state.   (\textbf{b}) Example brain regions identified as having low average controllabllity (\emph{pars opercularis}, blue), medium average controllability (post central, green) and high average controllability (isthmus cingulate, orange) (\textbf{c}) Simulation of example regions in panel \emph{(b)} differentially move the system into new functional states.  Stimulation applied to regions of high average controllability imparts more change in the functional brain state than stimulation applied to regions of low average controllability.
\label{fig4}}
\end{figure}

The results of systematically stimulating each brain region are shown in Fig.~\ref{fig5}.  We quantify the overall change in brain state configuration by measuring the functional effect of regional stimulation: the absolute value of the pairwise change in functional connectivity, averaged over all brain region pairs.  We observe that stimulation of regions with a high average controllability produce a large functional effect, while stimulation of regions with a high modal controllability result in a low functional effect (Fig.~\ref{fig5}a-b).

\begin{figure}
\centerline{\includegraphics{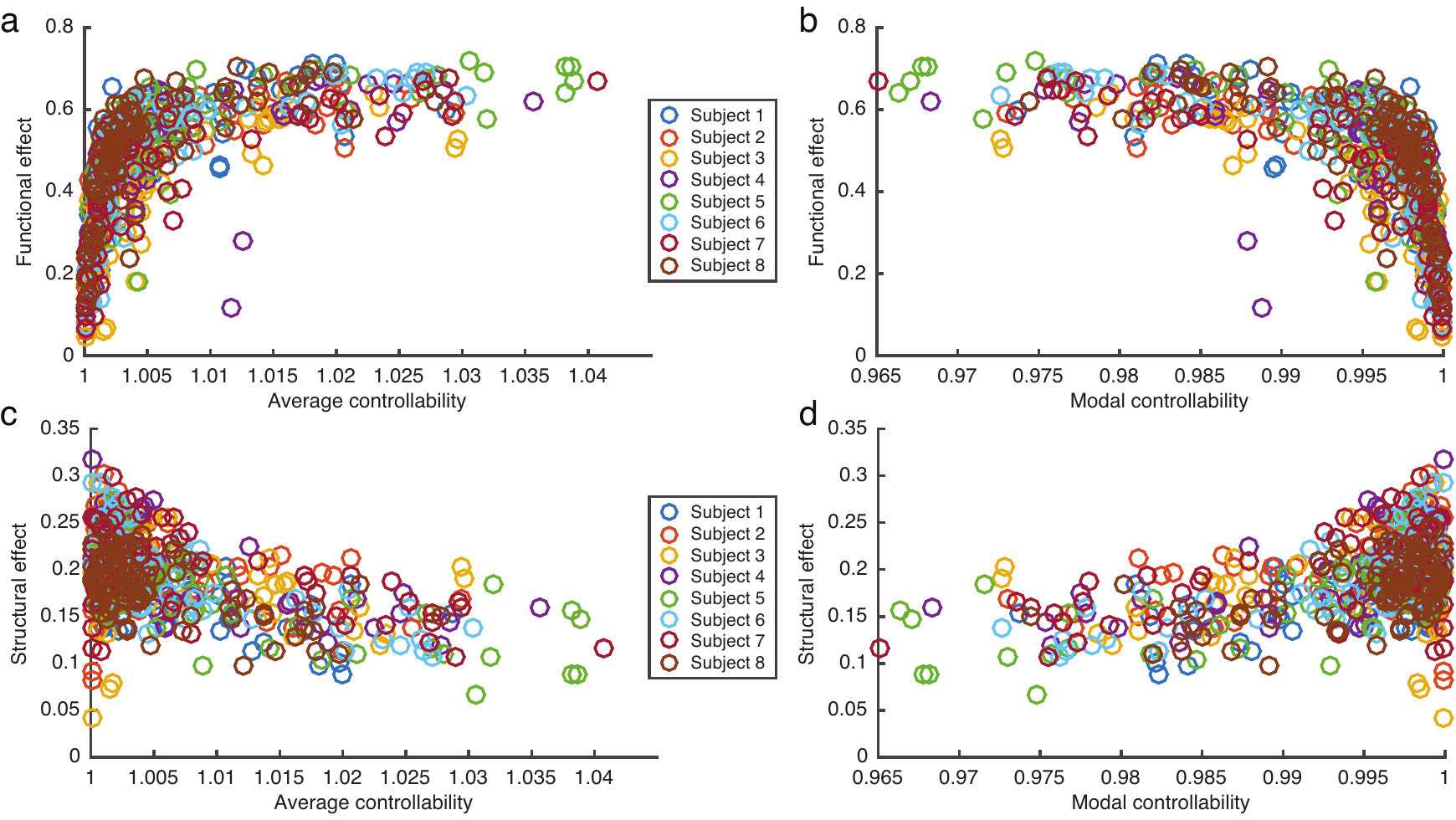}}
\caption{\textbf{Functional effect of stimulation.}  (\textbf{a-b}) The functional effect of regional stimulation plotted as a function of the average \emph{(a)} and modal \emph{(b)} controllability for each of the 83 brain regions.   (\textbf{c-d}) The structural effect of regional stimulation plotted as a function of the average \emph{(c)} and modal \emph{(d)} controllability for each of the 83 brain regions. Controllability predictions, simulations, and calculation of the functional and structural effects were performed for each of the three scans for each subject and the data points reflect values averaged over scans.
\label{fig5}}
\end{figure}

\subsection*{Structural connectivity differentially constrains the effects of regional stimulation}
We next ask how the underlying structural connectivity differentially constrains the functional effect of the stimulation for different brain regions.  We quantify this structural constraint by calculating the structural effect on network dynamics, which measures the change in spatial correlation between the structural connectivity matrix and the functional brain state matrix before and during stimulation (Methods).  Brain regions with a higher structural effect show a greater increase in similarity between structural and functional matrices when stimulated than regions with a lower structural effect.  Thus, stimulation of these regions is more constrained by the underlying structural connectivity.  As seen in Fig.~\ref{fig5}(c-d), the relationship between the structural effect and regional controllability is opposite that of the functional effect and controllability.  Stimulation of regions with a high average controllability results in a smaller structural effect, while stimulation of regions with a high modal controllability results in a higher structural effect.

In order to better understand why stimulation of regions with a high average controllability easily moved the system (high functional effect) and were less constrained by the underlying structure (low structural effect), we quantified the spread of activation from the regional stimulation.  Specifically, we asked if the stimulation of a given region induced focal or global changes in the brain state by calculating the fractional activation of the functional connectivity matrix.  The fractional activation is given by the fraction of pairwise regions that experience a change in their functional connectivity value that is above a given threshold (red pixels in Fig.~\ref{fig6}a-b).  If the stimulation of a brain region results in large changes that occur globally throughout the brain, the fractional activation will be high, but if the stimulation has only a focal effect, the fractional activation will be low.  

As one might expect, we observe a strong positive correlation between the functional effect and the fractional activation (Fig.~\ref{fig6}c, Spearman's $\rho=.992$, $p\ll.001$).  Specifically, a large functional effect is due to a global effect of regional stimulation that pushes the system into a nearby state, while a small functional effect is due to a focal effect of regional stimulation that moves the system toward a more distant state.  However, the relationship between the structural effect and fractional activation is more complex as seen in the crescent shaped curve of Fig.~\ref{fig6}d.  A maximum structural effect occurs at the curve of the crescent (arrow in Fig.~\ref{fig6}d), where the effects of stimulation are neither focal or global.  This result indicates that the underlying structural connections constrain the effects of stimulation the most in situations when regional stimulation impacts a moderately sized portion of the brain.

\begin{figure}
\centerline{\includegraphics{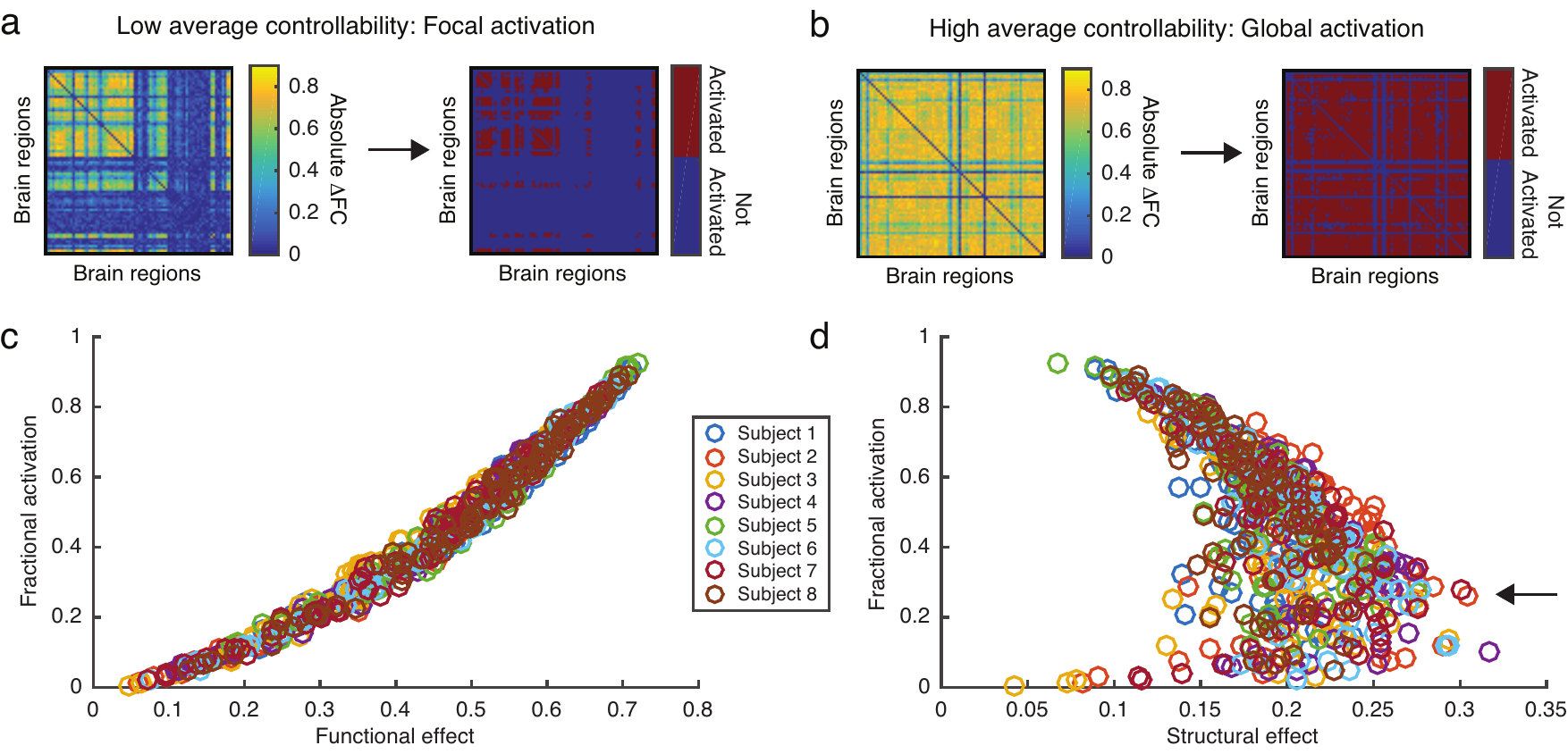}}
\caption{\textbf{Fractional activation.}  (\textbf{a-b}) The absolute change in functional connectivity and resulting fractional activation shown for a threshold value of 0.6. \emph{(a)} Example of stimulation applied to a region of low average controllability resulting in a focal effect on the resulting functional connectivity matrix.  \emph{(b)} Example of stimulation applied to a region of high average controllability resulting in a global effect on the resulting functional connectivity matrix. (\textbf{c})  The relationship between functional effect and fractional activation due to regional stimulation.  We observe a high positive correlation between functional effect and fractional activation (Spearman's $\rho=.992$, $p\ll.001$), indicating that a high functional effect corresponds to a global impact of stimulation while a low functional effect corresponds to a focal impact of stimulation. (\textbf{d}) The relationship between structural effect and fractional activation due to regional stimulation.  Calculations of the functional and structural effects and fractional activation were performed for each of the three scans for each subject and the data points reflect values averaged over scans.
\label{fig6}}
\end{figure}

\subsection*{Cognitive systems display varied roles in the structure-function landscape}
We next investigated the interplay between the structural and functional effects by examining the structure-function landscape (Fig.~\ref{fig7}).  Stimulation of individual brain regions revealed a range of tradeoffs between structural and functional effect values, with some regions displaying a high functional effect but low structural effect and others displaying a high structural effect but moderate or low functional effect.  We therefore asked if there was a relationship between the cognitive function associated with a brain region and its location in the structure-function landscape.  Brain regions were assigned to one of nine cognitive systems (Fig.~\ref{fig7} and Supplementary Material), eight of which were based on a data-driven clustering of functional brain networks \cite{Power:2011jm} that group regions that perform similar roles across a diverse set of tasks, as well as a group of subcortical regions. Although in reality no region has a singular function, we use these system assignments as a pragmatic means to assess whether controllability diagnostics are differentially identified in distributed brain networks.

In Fig.~\ref{fig7}a, we can see that some cognitive systems cover a wide range of the structure-function landscape, whereas other systems tend to be more localized.  Although the subcortical regions are well connected structurally, stimulation to any single subcortical region produces a functional effect that is less constrained by the underlying structural connections (low structural effect).  In contrast, regions in other systems remain similarly constrained by the underlying structural connections, but vary in their ability to impart a functional effect on the system.  Clusters of these nine cognitive systems emerge, suggesting that the parcellation scheme may be too fine-grained to understand the general organizing principles across the structure-function landscape; for example, there may be general network principles for sensory processing that are independent of modality (sight, sound, touch). Consequently, we created a more coarse-grained grouping of four cognitive systems (three functional and one structural) to examine broad-stroke differences among well-studied cognitive systems: the sensorimotor cortex, higher order cognitive, medial default mode network, and subcortical regions.  As seen in Fig.~\ref{fig7}b, regions in the sensorimotor cortex and higher order cognitive regions show a wide variation in their ability to impart a large functional effect when stimulated.  Interestingly, although regions in the default mode network are similarly constrained by structure when compared to the sensorimotor and higher order cognitive regions, regions within the default mode network consistently impart a large functional effect on the system.  Stimulation of subcortical regions also results in a large functional effect, but these regions are less constrained by structure than those in the default mode and therefore these two systems occupy different spaces in the structure-function landscape (Fig.~\ref{fig7}b; two sample, two-dimensional Kolmogorov-Smirnov test, $p=0.0003$).  This separation in the structure-function landscape could reflect the fact that the default mode network represents a functionally defined system, whereas subcortical regions are functionally diverse despite being well connected structurally.

In our final analysis, we examined why the default mode and subcortical regions had a stronger functional effect, and interestingly, why the subcortical structures showed a lower structural effect. We hypothesized that differential properties of these four systems can be better understood by investigating the density of connections between brain regions within a system versus the density of connections between systems.  Both the sensorimotor cortex and the higher order regions are defined functionally and are composed of multiple distributed structural systems.  They therefore, as a whole, have a low density of structural connections both within the system and between the system and the rest of the brain (Fig.~\ref{fig7}c).  In contrast, subcortical regions form a highly connected structural subnetwork while remaining well connected to the rest of the brain.  Thus, stimulation to these regions is globally distributed, resulting in a high functional effect.  The medial default mode network also forms a well-connected subnetwork, but is less connected to the sensorimotor cortex and higher order cognitive regions than the subcortical regions (see Table \ref{tab1}).  While stimulation to regions in the default mode network results in a large functional effect, regions in this subnetwork remain more constrained by the underlying structure. This latter result is consistent with previous work showing that brain regions in the default mode network display the highest correlation between structural and functional connectivity \cite{Horn:2014hu}. These results capture a mechanism for the subcortical regions to strongly influence regions across cerebral cortex where the default mode has a strong but targeted functional effect within its network, which enables these regions to quickly adapt from rest to a wide variety of task states \cite{Gu:2015dr}.

\begin{table}
\begin{center}
\small
\begin{tabular}{|l|c|c|c|c|}
\hline
 & Sensorimotor & Higher Order Cognitive & Default Mode & Subcortical \\
\hline
Sensorimotor & 0.054 & 0.025 & 0.059 & 0.094 \\
\hline
Higher Order Cognitive & 0.025 & 0.067 & 0.046 & 0.068 \\
\hline
Default Mode & 0.059 & 0.046 & 0.257 & 0.137 \\
\hline
Subcortical & 0.094 & 0.068 & 0.137 & 0.412 \\
\hline
\end{tabular}
\end{center}
\caption{Average density of connections between and within subnetworks of four cognitive systems.}
\label{tab1}
\end{table}

\begin{figure}
\centerline{\includegraphics[height=6in]{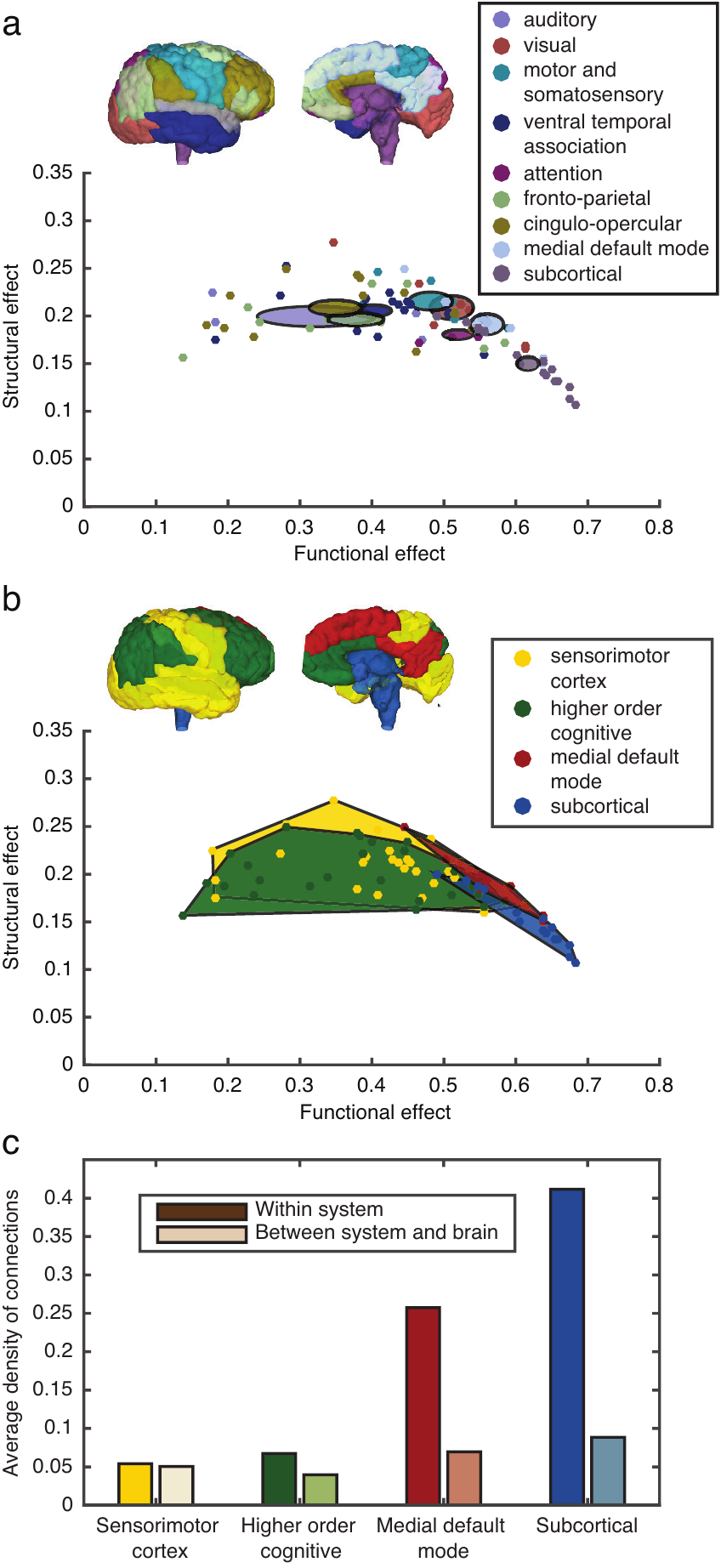}}
\caption{\textbf{Structure-function landscape.}  (\textbf{a})  Structural and functional effect values for stimulation of individual brain regions sorted into 9 cognitive systems.  Colored ellipses are centered on the mean structural and functional effects for a given cognitive system and the major and minor axis of the ellipse represent the standard error of the mean for the associated system. (\textbf{b}) Same as \emph{(a)} but data is further course grained into 4 cognitive system types as in \cite{Gu:2015fw}.  The colored regions indicate the convex hull surrounding the data points associated with the given system.  Simulations, and calculations were performed for each of the three scans for each subject and the data points reflect values averaged over both scans and individuals. (\textbf{c}) Average density of connections within and between the four cognitive system types.  Colors correspond to system assignments in \emph{(b)} and dark shades represent the average density of connections between regions within a single cognitive system while light shades represent the average density of connections between regions within that system and regions outside of the system.
\label{fig7}}
\end{figure}

\end{results}

\begin{discussion}
As neuromodulation is increasingly used to treat neurological disorders, it is essential to develop an understanding of the network-wide effects of focal stimulation. Such knowledge would directly inform the development of targeted protocols that effectively and efficiently maximize therapeutic benefits while minimizing the potential for adverse effects on brain dynamics and cognition.  An initial step towards achieving this goal lies in the examination of neuroimaging data through the lens of linear network control theory, a mathematical framework that predicts highly controllable brain regions from the pattern of underlying structural connectivity.  However, these techniques rely on the assumption that brain dynamics are linear, when in reality they are highly nonlinear.  

Here, we developed a computational modeling approach to investigate whether the predictions of a region's controllability drawn from a linear model could be validated in a nonlinear model. Specifically, we built data-derived structural connectivity matrices to computationally explore the effects of regional stimulation on network dynamics and functional states.  Using this model, we confirmed predictions from linear control theory, showing that stimulation of high average controllability regions resulted in global activation that produced little change in the topology of the functional connectivity, underscoring their role in moving the brain to nearby states.  Furthermore, we investigated the interplay between functional and structural effects of stimulation by examining how the global functional activity across brain regions was modulated by region-specific stimulation (a region's functional effect) and whether the region's structural connectivity accounted for its influence on the larger brain network (structural effect).  We observed that the underlying network connectivity differentially constrained the effects of stimulation: regions of high average controllability (strongly connected hubs \cite{Gu:2015dr}) displayed a high functional effect  -- meaning that they greatly increased the magnitude of functional connectivity -- while regions of low average controllability (weakly connected areas \cite{Gu:2015dr}) did not. Yet, stimulation that led to larger changes in functional connectivity magnitude (functional effect) induced global changes in functional connectivity topology (fractional activation), moving the system towards easily reachable states as opposed to the more distant states accessed through focal activation due to stimulation of low controllability regions.  Interestingly, when we parsed brain regions into cognitive systems, we found that stimulation of the medial default mode network showed both high structural and functional effects, differentiating it from other subnetworks by its ability to move the system while remaining influenced by the underlying network connectivity.

Perhaps one of the most striking observations from these data lies in the tradeoff between two competing consequences of stimulation: the magnitude of changes in functional connectivity, and the spatial specificity of changes in functional connectivity. We observed that stimulation to network hubs, predominantly located in default mode and subcortical structures \cite{Hagmann:2008gd,Bassett:2011bm,Gu:2015dr}, induces widespread increases in the magnitude of functional connectivity between brain regions. Yet, this broad impact is affected only at the expense of spatial specificity. In contrast, stimulation targeted to weakly connected areas (low average controllability) predominantly located in fronto-parietal regions \cite{Gu:2015dr}, induces focal changes in functional connectivity. The differential impact of stimulation to these two strongly \emph{versus} weakly connected regions suggests the possibility of two different classes of therapeautic interventions: (i) a broad reset, in which brain dynamics are globally altered, and (ii) a focal change, in which brain dynamics of a few regions are altered.  These differential outcomes may offer mechanistic insights into the role of stimulation in distinguishing fine-scale differences between the concepts \cite{Lecce2015} \emph{versus} broadly altering general cognitive processes \cite{Kraft2015,Bonni2015} or brain states \cite{Crescentini2015}.

The differences in the impact of stimulation to hubs \emph{versus} non-hubs further supports the predictions of linear network control theory \cite{Pasqualetti2014} and their recent applications to neuroimaging data \cite{Gu:2015dr}. In this prior work, theoretical insights from structural controllability \cite{Kailath1980} were used to make the prediction that network hubs -- particularly in the default mode system -- facilitate the movement of the brain to many easily reachable states. In contrast, weakly connected nodes of the network -- particularly in cognitive control systems -- were predicted to facilitate the movement of the brain to difficult-to-reach states. Our results confirm these predictions and offer further insights into the mechanisms of these control strategies. Specifically, easily reachable states are those that display patterns of functional connectivity that are very similar to those observed in the initial state, while difficult-to-reach states are those that display patterns of functional connectivity that are very different from those observed in the initial state.

In addition to these large-scale observations of network state change, we also probed the degree to which the pattern of functional connectivity (whether focal or global) was correlated with the pattern of structural connectivity. Prior work has focused largely on the relationship between resting state functional connectivity and structural connectivity \cite{Honey2009,Honey2010,Goni2014}, under the assumption that the brain's resting baseline might be highly constrained by anatomy. However, structural connections likely constrain functional connectivity present in all brain states, irrespective of the cognitive process at play \cite{Gu:2015dr}. Indeed, a few recent studies have demonstrated the non-trivial relationships between individual differences in the pattern of structural connections and the observed functional connectivity across multiple cognitive states \cite{Hermundstad:2013dm,Hermundstad:2014bg}. Here we observe that the similarity between structural connectivity and observed functional connectivity depends significantly on the brain region that was stimulated. Critically, this relationship was not driven by the average controllability of the region. Along with prior links between average controllability and degree \cite{Gu:2015dr}, these results suggest that structural constraints on stimulation-elicited functional connectivity can not easily be predicted by whether a region is a hub or a non-hub. 

The relative independence of the functional and structural effects is particularly evident across large-scale cognitive systems. Indeed, we observe an inverted U-shaped curve between these two variables: systems that display a middling change in the magnitude of functional connectivity with stimulation tend to display functional connectivity patterns that are most reminiscent of the underlying structural connectivity. In contrast, systems that display a very large or very small change in the magnitude of functional connectivity with stimulation tend to display functional connectivity patterns that are very different from the structural connectivity. Intuitively, stimulation to hubs or non-hubs produces brain states that are far from those simply predicted by structural connectivity, and potentially thus far from normative \cite{Hermundstad:2014bg}. It will be interesting in future to determine which brain states elicited by stimulation are consistent \emph{versus} inconsistent with states observed in normative brain dynamics. Such a question is reminiscent of similar work in control theory identifying so-called \emph{allowable} transitions \cite{Cornelius2013,Sun2013}. More broadly, an understanding of potential brain states elicited by stimulation is key to deploying stimulation in such a way as to maximize clinical benefit while minimizing pathological configurations of the network.

There are several important methodological considerations pertinent to this work. While this work represents an important step in characterizing the effects of stimulation and control in nonlinear brain networks, it should be noted that the nonlinear model of brain dynamics employed here, while biologically inspired, is a simplified mean-field model of neuronal dynamics.  Additionally, we have performed only a rough partitioning of the brain into regions and finer scales of regional partitioning could lead to greater distinction between regional roles as more subtle patterns of brain connectivity are revealed.  Future work is necessary to confirm the effects of spatial resolution on models of targeted stimulation.

Finally, an important finding of this study was that while we observed a high level of reproducibility in simulations run using connectivities derived from separate scans within a single subject, we observed variance across subjects in the range of coupling corresponding to the fixed point and oscillatory regimes of the model.  While measures of controllability, functional, and structural effects collapsed to the same curves across subjects (Fig.~\ref{fig5}), the shifting of the oscillatory regime in the coupling parameter space (Fig.~\ref{fig2}c) indicates that the model is sensitive to variances in structural connectivities between individuals.  Although the set of 8 subjects studied here is underpowered to study individual variation in connectivity and its impact on model performance, this encouraging result suggests the utility of such approaches to understand individual variability in structural connections and how it may change the functional effect of stimulation. This modeling approach provides a means to individualize stimulation protocols for personalized medical treatments or performance enhancements.

\end{discussion}

\begin{methods}

\subsection*{Human DSI data acquisition and preprocessing}
Diffusion spectrum images (DSI) were acquired from a total of $8$ subjects in triplicate (mean age $27\pm5$ years, $2$ female, $2$ left handed) along with a $T1$ weighted anatomical scan at each scanning session \cite{Cieslak:2014ic}. DSI scans sampled $257$ directions using a $Q5$ half shell acquisition scheme with a maximum $b$ value of $5000$ and an isotropic voxel size of $2.4$mm. We utilized an axial acquisition with the following parameters: $TR=11.4$s, $TE=138$ms, $51$ slices, FoV ($231$,$231$,$123$ mm). All participants volunteered with informed consent in accordance with the Institutional Review Board/Human Subjects Committee, University of California, Santa Barbara.

DSI data were reconstructed in DSI Studio (www.dsi-studio.labsolver.org) using $q$-space diffeomorphic reconstruction (QSDR) \cite{Yeh:2011ca}. QSDR first reconstructs diffusion weighted images in native space and computes the quantitative anisotropy (QA) in each voxel. These QA values are used to warp the brain to a template QA volume in MNI space using the SPM nonlinear registration algorithm. Once in MNI space, spin density functions were again reconstructed with a mean diffusion distance of $1.25$ mm using three fiber orientations per voxel. Fiber tracking was performed in DSI Studio with an angular cutoff of $55^{\circ}$, step size of $1.0$ mm, minimum length of $10$ mm, spin density function smoothing of $0.0$, maximum length of $400$ mm and a QA threshold determined by DWI signal in the CSF. Deterministic fiber tracking using a modified FACT algorithm was performed until $100,000$ streamlines were reconstructed for each individual.

Anatomical scans were segmented using FreeSurfer \cite{Dale:1999ks} and parcellated according to the Lausanne 2008 atlas included in the connectome mapping toolkit \cite{Hagmann:2008gd}. A parcellation scheme including $83$ regions was registered to the B0 volume from each subject's DSI data. The B0 to MNI voxel mapping produced via QSDR was used to map region labels from native space to MNI coordinates. To extend region labels through the gray/white matter interface, the atlas was dilated by $4$mm. Dilation was accomplished by filling non-labeled voxels with the statistical mode of their neighbors' labels. In the event of a tie, one of the modes was arbitrarily selected. Each streamline was labeled according to its terminal region pair.

\subsection*{Construction of structural brain networks}
We define structural brain networks by subdividing the entire brain into 83 anatomically distinct brain areas (network nodes) \cite{Cammoun:2012gk}. Consistent with prior work \cite{Bassett:2011bm,Hermundstad:2013dm,Hermundstad:2014bg}, we connect nodes by the number of white matter streamlines identified by a commonly used deterministic tractography algorithm described above \cite{Cieslak:2014ic} and normalized by the sum of the volumes of the nodes. This procedure results in sparse, weighted, undirected structural brain networks for each subject ($N=8$) and each scanning session ($n=3$), where network connections represent the density of white matter tracts between brain regions. The definition of structural brain networks based on tractography data in humans follows from our primary hypothesis that control features of neural dynamics are in part determined by the structural organization of the brain's white matter tracts.

\subsection*{Mathematical model of brain dynamics}
We employ a data-driven, nonlinear model of brain dynamics that allows for the systematic study of the effects of stimulation to different nodes in the network.  Brain regions (modeled by a single network node) are governed by Wilson-Cowan equations \cite{Wilson:1972kj} representing the population-level activity of the $j^{th}$ region as follows:
\begin{eqnarray*}
\tau\frac{dE_j}{dt}&=&-E_j(t)+\left(S_{e\_max}-E_j(t)\right)S_e\left(c_1E_j(t)-c_2I_j(t)+c_5\sum_k{A_{jk}E_{k}(t-\tau_d^k)}+P_j(t)\right)+\sigma w_j(t)\\
\tau\frac{dI_j}{dt}&=&-I_j(t)+\left(S_{i\_max}-I_j(t)\right)S_i\left(c_3E_j(t)-c_4I_j(t)\right)+\sigma v_j(t)
\end{eqnarray*}
where $E(t)/I(t)$ represent the firing rate of the excitatory/inhibitory population respectively, $\tau=8$ ms is a time constant, $P(t)$ is an external stimulus parameter, and $w_j(t)$ and $v_j(t)$ are drawn from a standard normal distribution and act as additive noise to the system with $\sigma=0.00001$.  Regions are coupled through the excitatory population with a connectivity matrix $\bf{A}$ derived from a single scan of an individual subject's tractography data, parcellated to give a total of 83 brain regions, and normalized by region size.  Delays, $\tau_d$, between regions are calculated as a function of physical distance between identified brain regions, assuming a transmission velocity of $10$ m/s (range $\tau_d=0.8-14.8$ ms). The transfer function is given by the sigmoidal function
\begin{equation*}
S_{e/i}(x)=\frac{1}{1+e^{(-a_{e/i}(x-\theta_{e/i}))}}-\frac{1}{1+e^{a_{e/i}\theta_{e/i}}}.
\end{equation*}
Model constants are set as in \cite{Wilson:1972kj} to be $c_1=16$, $c_2=12$, $c_3=15$, $c_4=3$, $a_e=1.3$, $a_i=2$, $\theta_e=4$, $\theta_i=3.7$.  These parameter choices imply that an uncoupled oscillator has three states: a low fixed point, limit cycle, and high fixed point.  For a single oscillator, increasing the current input into the oscillator, $P(t)$, will allow it to transition between the three states.  In our model of coupled oscillators, the global coupling parameter, $c_5$, serves as an additional mechanism for allowing the system to transition between these states for fixed values of $P(t)$.   To simulate brain activity, we set $P(t)=0$ for all regions, and we model stimulation to single brain region, $j$, by setting $P_j(t)=1.25$. This value of $P$ implies that an uncoupled oscillator will enter into limit cycle activity with a frequency near 20 Hz which is inline with the biological range for oscillatory brain activity. All simulations are allowed to initially stabilize for $1$s before analysis, and the temporal dynamics of the $j^{th}$ brain region are given by the firing rate of the excitatory population, $E_j(t)$.

\subsection*{Evaluating oscillatory transition parameters}
In order to asses the point in the parameter space at which the system switched from the low fixed point to the oscillatory regime, we ran 1s simulations in which no stimulating current was applied ($P=0$) for increasing values of the global coupling parameter ($c_5=1.0$ to $c_5=1.5$ in step sizes of 0.05).  The average firing rate of the excitatory population for each region was recorded as a function of the global coupling parameter and is shown for stimulations using a single scan from two separate subjects in Fig.~S2.  At a certain value of the global coupling parameter that varied between scans and subjects, we see a sudden increase in the firing rate across most regions, indicating the transition of the system from the fixed point to the oscillatory regime.  We use the value of $c_5$ at which this transition occurs to assess the reproducibility and variability within and between subjects.  For all simulations assessing the impact of regional stimulation, the value of $c_5$ immediately before this transition was used.

\subsection*{Inter- and intra-subject reproducibility and variability}
Inter- and intra-subject reproducibility were calculated using the intraclass correlation coefficient (ICC), and assuming a random effect model as in \cite{Wei:2004hg}. The between subject ICC is estimated using the mean squares (MS) obtained by applying ANOVA with the subject as the factor to the value of global coupling at which the system switches to the oscillator state obtained as described above:
\begin{equation}
ICC_B=\frac{J(SMS-EMS)}{J*SMS+I*RMS+(IJ-I-J)EMS}.
\end{equation}
Similarly, the within subject ICC [jmv6]is given by
\begin{equation}
ICC_W=\frac{I(RMS-EMS)}{J*SMS+I*RMS+(IJ-I-J)EMS},
\end{equation}
where $I$ is the number of subjects, $J$ is the number of scans, $SMS$ is the MS between scans, $RMS$ is the MS between subjects, and $EMS$ is the MS due to error.

The between subject variability was defined to be the variance of the mean value of the transition point obtained for each subject, and the within subject variability was defined to be the within subject variance, averaged across all subjects.

\subsection*{Linear network control theory}
To study the theoretical ability of a certain brain region to influence other regions in arbitrary ways we adopt the control theoretic notion of \emph{controllability}. Controllability of a dynamical system refers to the possibility of driving the state of a dynamical system to a specific target state by means of an external control input \cite{Kalman:1963wk}.  As in \cite{Gu:2015dr}, we employ a simplified noise-free linear discrete-time and time-invariant network model:
\begin{equation}\label{eq: linear network}
  \mathbf{x} (t+1) = \mathbf{A} \mathbf{x}(t) + \mathbf{B}_{\mc K} \mathbf{u}_{\mc K} (t) ,
\end{equation}
where $\mathbf{x}$ describes the state of brain regions over time, $\mathbf{A}$ is a normalized structural connectivity matrix derived from tractography data as described above. The Supplementary Material contains a discussion of the method used to normalize matrices across subject data sets.  The input matrix $\mathbf{B}_{\mc K}$ identifies the control points, and $\mathbf{u}_{\mc K}$ denotes the control strategy.

Classic results in control theory ensure that controllability of the network \eqref{eq: linear network} from the set of network nodes $\mc K$ is equivalent to the controllability Gramian $\mathbf{W}_{\mc K}$ being invertible, where
\begin{equation}
\mathbf{W}_{\mathcal{K}} = \sum_{\tau =0}^{\infty}\mathbf{A}^\tau
\mathbf{B}_{\mathcal{K}}\mathbf{B}_{\mathcal{K}}^\transpose \mathbf{A}^\tau.
\end{equation}
We utilize this framework to choose control nodes one at a time, and thus the input matrix $B$ in fact reduces to a one-dimensional vector.

We examine 2 diagnostics of controllability utilized in the network control literature: \emph{average controllability} and \emph{modal controllability} \cite{Gu:2015dr}.  Average controllability of a network equals the average input energy from a set of control nodes and over all possible target states \cite{Marx:2004tn,Shaker:2012kq}. As in \cite{Gu:2015dr}, we adopt $\text{Trace}(W_K)$ as a measure of average controllability.  Regions with high average controllability are, on average, most influential in the control of network dynamics over all nearby target states with least energy.  Modal controllability refers to the ability of a node to control each evolutionary mode of a dynamical network \cite{HAMDAN:1989gm}, and can be used to identify states that are difficult to control from a set of control nodes. Modal controllability is computed from the eigenvector matrix $V = [v_{ij}]$ of the network adjacency matrix $\mathbf{A}$. Regions with high modal controllability are able to control all the dynamic modes of the network, and hence they can drive the dynamics towards hard-to-reach configurations.

The stimulation paradigm that we study is a constant input over a short period of time, whereas the controllability metrics discussed above assume a more general time-varying input. We treat this constant current stimulation as a a simple approximation of the more general stimulation paradigm traditionally studied in the control literature, and therefore use average and modal controllability to assess control architecture. Because our stimulation paradigm is given by a constant input, one can also examine the steady state response of the network.  In the Supplementary Information and Fig.~S1, we study the more constrained steady state response of the network, and demonstrate that our approximation presented here is an accurate one.  However, it should be noted that the statistics we use here (average/modal controllability) may not always be an adequate approximation of the steady state response of a network, and therefore care should be taken in using these statistics in other studies without first demonstrating their consistency with the steady state response statistics.

 \subsection*{Quantifying functional brain states}
 We quantify brain states by calculating the pairwise maximum normalized cross-correlation \cite{Kramer:2009ba,Feldt:2007dw} between the firing rate of the excitatory populations, $E_i(t)$ and $E_j(t)$, for brain regions $i$ and $j$.  All calculations are performed using a $1$ s window and a maximum lag of $250$ ms.

\subsection*{Functional effect of stimulation}
Simulations of neural dynamics are first allowed to stabilize for $1$s to reach the stable activity from the global coupling parameter and then the remaining time is divided into two parts: a stimulation free period of $1$s, followed by a $1$s period of stimulation.   During the stimulation period, a single region, $s$, is selected and a stimulus is applied to this region by setting $P_s(t)=1.25$, while $P(t)=0$ for all other regions.  Functional brain states are computed separately for the stimulation-free (`before' in Fig.~\ref{fig4}) and stimulation (`during' in Fig.~\ref{fig4}) periods.  We assess the pairwise change in functional brain states by subtracting the correlation values obtained in the stimulation-free time window before the stimulation is applied from the correlation values obtained in the time window during the stimulation.  We then measure the average change in functional brain states, termed the \emph{functional effect}, as the absolute value of this difference averaged over all region pairs.  The greater the deviation of the functional effect from zero, the greater the effect of stimulation on brain state reconfiguration.

\subsection*{Structural effect on network dynamics}
In order to assess how the structural brain connectivity constrains network dynamics, we calculate the similarity, defined by the 2-dimensional correlation coefficient, between the structural connectivity matrix used as a basis of the simulation and the functional matrix describing the resultant brain state.  This calculation is done first using the initial functional brain state before stimulation, and then using the functional brain state during the stimulation period.  Stimulation of a single region increases the similarity between the structural connectivity and functional brain state, and we quantify this increase, termed the \emph{structural effect}, by subtracting the correlation obtained before stimulation from that obtained during stimulation.  Thus, a high structural effect indicates that the underlying connectivity structure constrains the functional effects of stimulation to this region.

\subsection*{Fractional activation}
The fractional activation due to the stimulation of a single brain region is defined to be the fraction of pairwise regions that experience a change in their functional connectivity value that is above a given threshold.  The change in functional connectivity is calculated as described above to calculate the \emph{functional effect} by assessing the absolute change in pairwise correlation values before and during regional stimulation.  If the stimulation of a brain region results in large changes that occur globally throughout the brain, the fractional activation will be high, but if the stimulation has only a focal effect, the fractional activation will be low.  Here, we report findings using a threshold value of 0.6, however, results were similar across a range of thresholds (Fig.~S3.)

\end{methods}

\references{WC_bib}

%

\section*{Acknowledgements}
Research was sponsored by the Army Research Laboratory and was accomplished under Cooperative Agreement Number W911NF-10-2-0022.  DSB acknowledges support from the John D. and Catherine T. MacArthur Foundation, the Alfred P. Sloan Foundation, the Army Research Office (W911NF-14-1-0679), the National Institute of Mental Health (2-R01-DC-009209-11), the National Institute of Child Health and Human Development (1R01HD086888-01), the Office of Naval Research, and the National Science Foundation (BCS-1441502 and BCS-1430087).  FP acknowledges support from the National Science Foundation (BCS-1430279) and STG acknowledges support from the Institute for Collaborative Biotechnologies through the US Army Research Office (W911NF-09-0001).  The content is solely the responsibility of the authors and does not necessarily represent the official views of any of the funding agencies.

\section*{Author Contributions}
SFM, FP, JV, and DSB conceptualized the project; SFM, SG, MC, and STG contributed code/data; SFM analyzed the data; and SFM, FP, JV, and DSB wrote the manuscript.

\end{document}


\maketitle
%
%

\newpage
\clearpage

\section*{Normalization of structural matrices across subjects}
When doing controllability calculations, we must ensure that the structural adjacency matrix describing the tractography connectivity is Schur stable.  In Gu et al.\cite{Gu:2015dr}, this was obtained by normalizing each structural matrix by one plus its largest singular value.  Because each matrix effectively received a slightly different normalization factor, regional controllability results were ranked in order to perform averaging across scans and subjects.  To avoid the loss of information that results from ranking the data, we instead employed a different form of normalization.  We first calculated the maximum eigenvalue for each structural matrix in the data set.  From this pool of maximum eigenvalues, we then selected the maximum value, and divided all structural matrixes by two times this quantity.  This ensures Schur stability and allows us to compare directly compare regional controllability values obtained from different structural matrices.  The controllability results presented in this paper therefore represent the resultant numerical values of regional controllability calculations, as opposed to the ranked values presented in Gu et al\cite{Gu:2015dr}.

\section*{Average controllability and the steady state response}
In linear network control theory, the controllability of a network refers to the possibility of altering the configuration of the network nodes via external stimuli and in a predictable way. To quantify the degree of controllability of a network, we first model the dynamical interaction among network nodes by means of a discrete-time, linear, time-invariant system:
\begin{align*}
  x(t+1) = A x(t) + B_K u(t).
\end{align*}
In the equation above, x is a vector containing the states of the network nodes, $A = A^\transpose$ is a (stable) weighted adjacency matrix of the network, $u$ is the external control signal, and
$B_K$ identifies the control nodes; see also \cite{Gu:2015dr,Pasqualetti2014}.

In our simulations, we use a constant input current to stimulate brain regions, and therefore are interested in the steady state response of the network.  With a constant control input, the network steady state is
\begin{align*}
  x_\text{steady} = (I - A)^{-1} B_{K} u_{\text{constant}},
\end{align*}
where $u_{\text{constant}}$ is the value of the constant input. Thus, the steady state effect of a constant input to the $i$-th region is characterized by the $i$-th column of the matrix $(I - A)^{-1}$. The largest value of the $i$-th column gives steady state response of the region maximally effected by the input, while the average of the $i$-th column gives the average steady state effect over all regions.

The use of a constant input as regional stimulation is a special case of the more general paradigm of a time-varying input normally used to define network control statstics such as those used in the main manuscript (average/modal controllability).  We therefore would like to relate this steady state response to the average controllability which describes the more general paradigm.   The degree of controllability of a network can be quantified in different ways  \cite{Pasqualetti2014}, but in this paper, we use the classical definition of the Controllability Gramian, that is,
\begin{align*}
  W_K = \sum_{\tau = 0}^\infty A^\tau B_K B_K^T A^\tau,
\end{align*}
and measure the \emph{average} degree of controllability as $\text{Trace}(W_K)$, which has a specific system theoretic interpretation \cite{Gu:2015dr,Pasqualetti2014,Kailath:1980wa}. Moreover, notice that
\begin{align*}
  \text{Trace} (W_K) &= \text{Trace} \left( \sum_{\tau = 0}^\infty A^\tau B_K
                       B_K^T A^\tau \right) \\
                     &= \sum_{\tau = 0}^\infty \text{Trace} \left( A^{2 \tau} B_K
                       B_K^\transpose \right) \\
                     &= \text{Trace} \left( \sum_{\tau = 0}^\infty
                       A^{2\tau} B_K  B_K^\transpose \right) \\
                     &= \text{Trace} \left( (I-A^2)^{-1} B_K
                       B_K^\transpose\right) \\
                     &= \sum_{i \in K} (I-A^2)^{-1}_{ii} .
\end{align*}
In other words, the controllability degree with control nodes $K$ equals the sum of the diagonal entries of $(I-A^2)^{-1}$ indexed by $K$. Because of the normalization of the adjacency matrix adopted
in this work, it can be verified that
\begin{align*}
  (I - A)^{-1} \approx (I-A^2)^{-1},
\end{align*}
so that the average controllability information can be reconstructed from the steady state response matrix $(I - A)^{-1}$.  Specifically, for stimulation of a single region, the largest entry of the $i$-th column of the steady state matrix will be approximately equal to the average controllability. 

As seen in Fig.~S1a, when we plot the functional effect of stimulation as a function of the largest steady state value, we do indeed reproduce the results of Fig. 5a.  Additionally, we see a similar result when plotting the functional effect of stimulation for the average steady state value (Fig.~S1b).  Therefore, in the main manuscript, we present our findings in terms of the more general regional controllability values instead of the steady state response, which also allows for comparison with previous work using these measures to study the properties of structural brain networks \cite{Gu:2015dr}.  However, it should be noted that the average/modal controllability may not always be an adequate approximation of the steady state response of a network, and therefore care should be taken in using these statistics in other studies without first demonstrating their consistency with the steady state response statistics.

\section*{Mapping regions to cognitive systems}
Similar to the assignment of brain regions in Gu et al.\cite{Gu:2015dr} and inspired by Power et al.\cite{Power:2011jm}, we initially assigned each of the 83 brain regions to one of 9 cognitive systems.  Our only divergence from Gu et al. was the creation of a ``ventral temporal association'' category to bin perceptual regions associated with invariant object representations and multisensory activation.  For further analysis, this assignment was coarse grained into 4 cognitive systems \cite{Gu:2015fw}.  The placement of each region into each cognitive system is summarized in Table S1.

\newpage

\begin{table}
\begin{center}
\small
\begin{tabular}{|l|l|l|}
\hline
\bf{Region Name} & \bf{9 system assignment} & \bf{4 system assignment} \\
\hline
Lateral Orbitofrontal & attention & higher order cognitive  \\
\hline
Pars Orbitalis & cingulo-opercular & higher order cognitive  \\
\hline
Frontal Pole & fronto-parietal & higher order cognitive  \\
\hline
Medial Orbitofrontal & fronto-parietal & higher order cognitive  \\
\hline
Pars Triangularis & fronto-parietal & higher order cognitive  \\
\hline
Pars Opercularis & cingulo-opercular & higher order cognitive  \\
\hline
Rostral Middle Frontal & cingulo-opercular & higher order cognitive  \\
\hline
Superior Frontal & medial default mode & medial default mode  \\
\hline
Caudal Middle Frontal & fronto-parietal & higher order cognitive  \\
\hline
Precentral  & motor and somatosensory & sensorimotor cortex  \\
\hline
Paracentral & motor and somatosensory & sensorimotor cortex  \\
\hline
Rostral Anterior Cingulate & cingulo-opercular & higher order cognitive  \\
\hline
Caudal Anterior Cingulate & cingulo-opercular & higher order cognitive  \\
\hline
Posterior Cingulate & medial default mode & medial default mode  \\
\hline
Isthmus Cingulate & medial default mode & medial default mode  \\
\hline
Post Central & motor and somatosensory & sensorimotor cortex  \\
\hline
Supramarginal & cingulo-opercular & higher order cognitive  \\
\hline
Superior Parietal & attention & higher order cognitive  \\
\hline
Inferior Parietal & fronto-parietal & higher order cognitive  \\
\hline
Precuneus & medial default mode & medial default mode  \\
\hline
Cuneus & visual & sensorimotor cortex  \\
\hline
Pericalcarine & visual & sensorimotor cortex  \\
\hline
Lateral Occipital & visual & sensorimotor cortex  \\
\hline
Lingual  & visual & sensorimotor cortex  \\
\hline
Fusiform & ventral temporal association & sensorimotor cortex  \\
\hline
Parahippocampal & ventral temporal association & sensorimotor cortex  \\
\hline
Entorhinal Cortex & ventral temporal association & sensorimotor cortex  \\
\hline
Temporal Pole & ventral temporal association & sensorimotor cortex  \\
\hline
Inferior Temporal & ventral temporal association & sensorimotor cortex  \\
\hline
Middle Temporal & ventral temporal association & sensorimotor cortex  \\
\hline
Bank of the Superior Temporal Sulcus & ventral temporal association & sensorimotor cortex  \\
\hline
Superior Temporal & auditory & sensorimotor cortex  \\
\hline
Transverse Temporal & auditory & sensorimotor cortex  \\
\hline
Insula & fronto-parietal & higher order cognitive  \\
\hline
Thalamus & subcortical & subcortical  \\
\hline
Caudate & subcortical & subcortical  \\
\hline
Putamen & subcortical & subcortical  \\
\hline
Pallidum & subcortical & subcortical  \\
\hline
Nucleus Accumbens & subcortical & subcortical  \\
\hline
Hippocampus & subcortical & subcortical  \\
\hline
Amygdala & subcortical & subcortical  \\
\hline
Brainstem & subcortical & subcortical  \\
\hline
\end{tabular}
\end{center}
\caption{Assignment of brain regions to cognitive systems.}
\label{tabS1}
\end{table}

\newpage

\begin{figure}
\centerline{\includegraphics{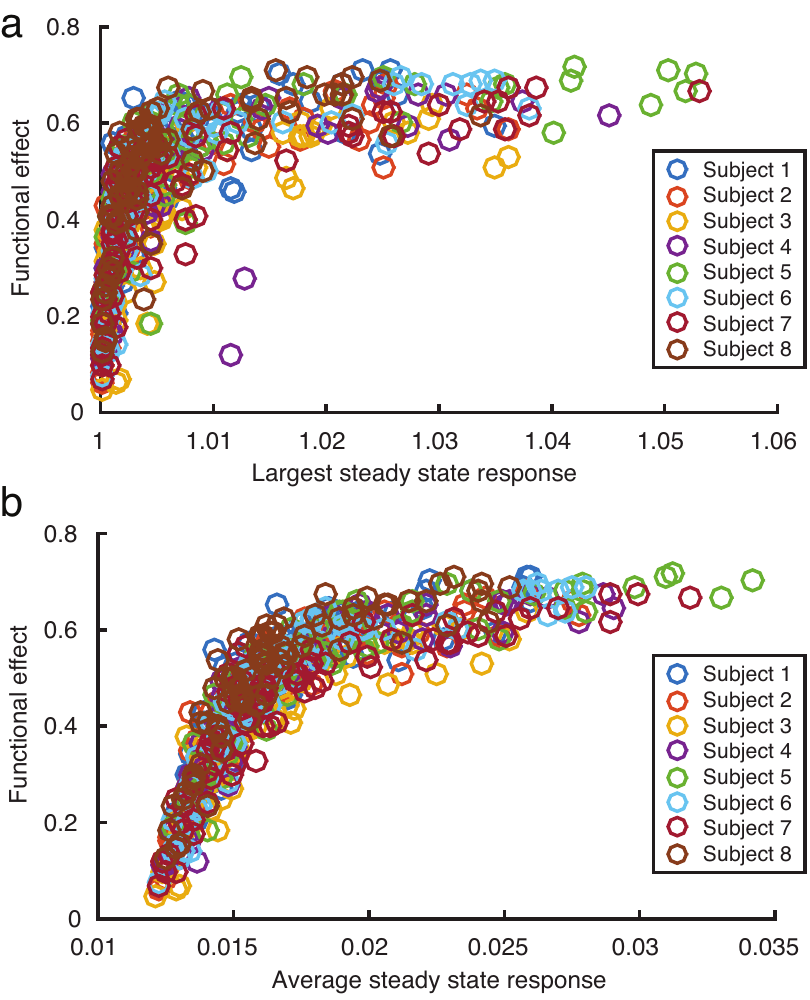}}
\caption{\textbf{Steady state response and functional effect.}   The functional effect resulting from regional stimulation plotted as a function of (\textbf{a}) the largest steady state value (\textbf{b}) the average steady state value.
\label{figS1}}
\end{figure}

\begin{figure}
\centerline{\includegraphics{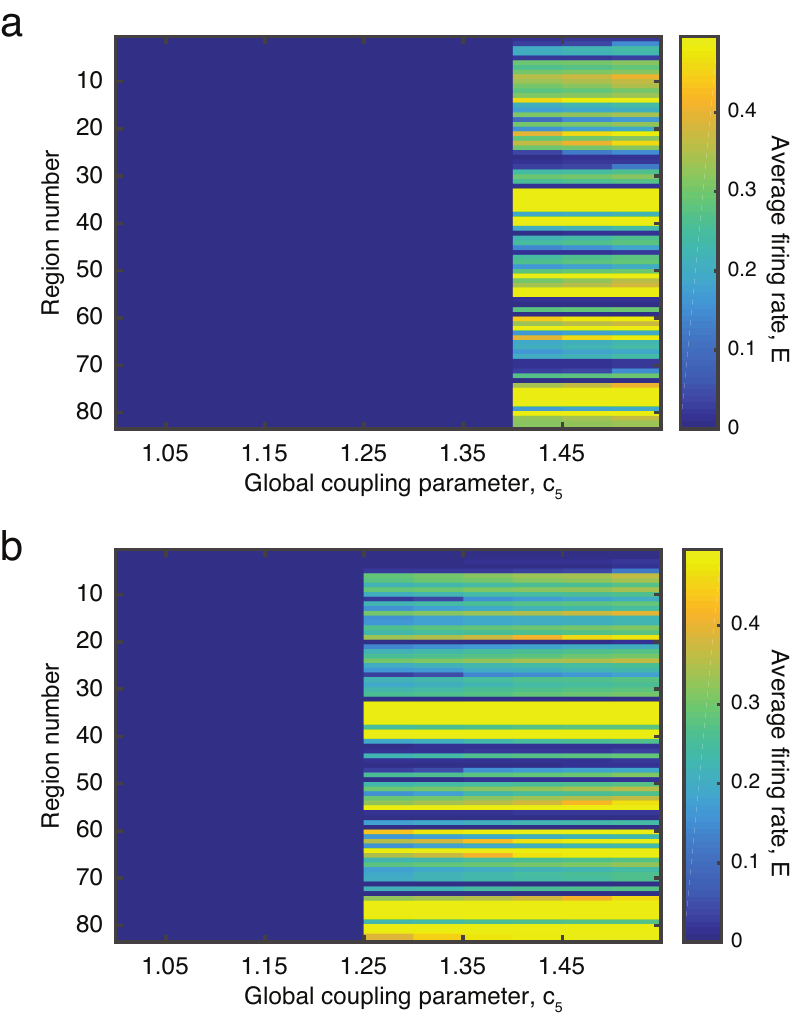}}
\caption{\textbf{Transition to oscillatory regime.}  (\textbf{a-b})  Examples of the transition to the oscillatory regime in simulations from a single scan obtained from two different subjects.  In (a) the transition occurs for a global coupling value of $c_5=1.4$ whereas in (b) the transition occurs for $c_5=1.25$.
\label{figS2}}
\end{figure}

\begin{figure}
\centerline{\includegraphics{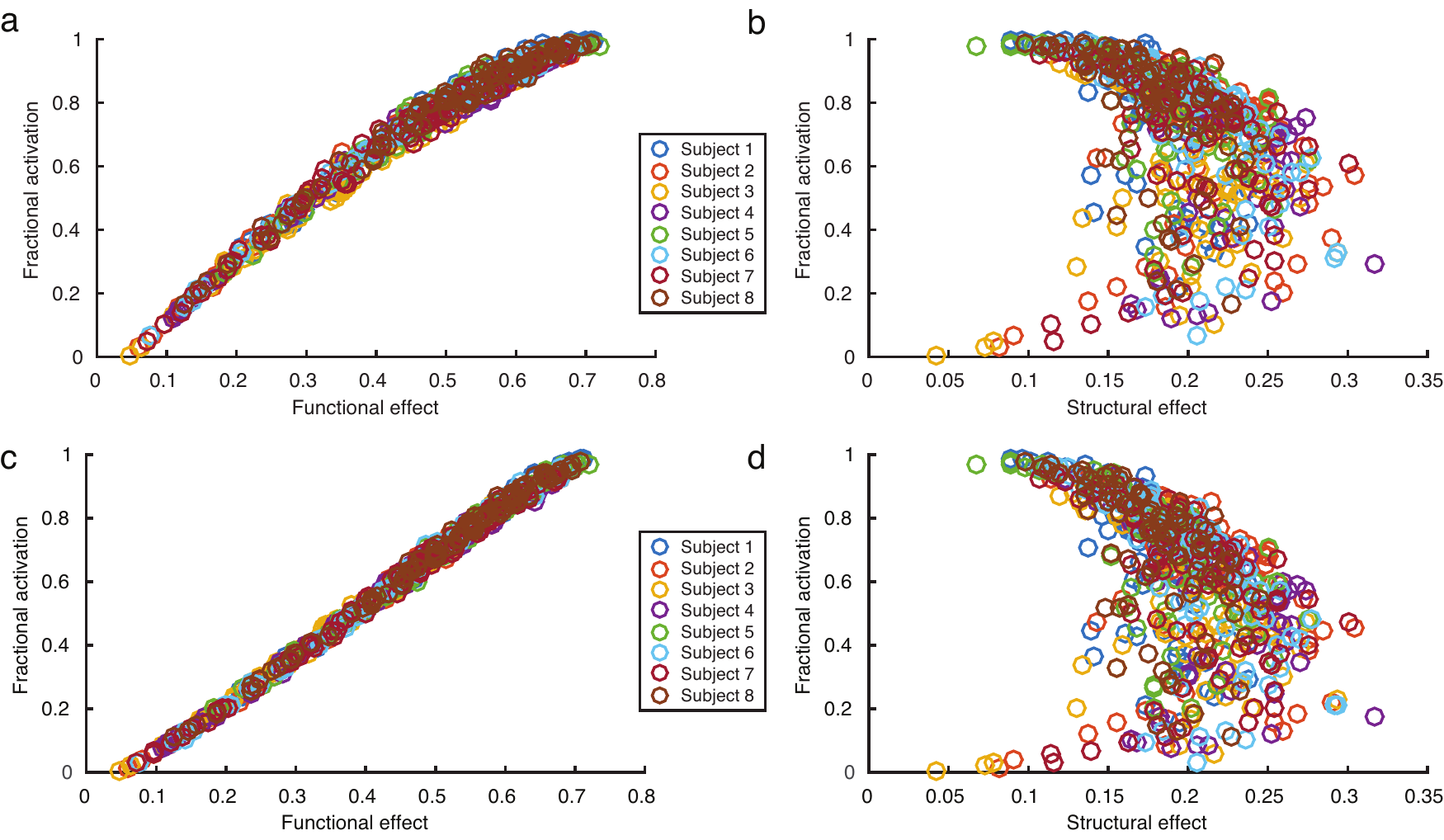}}
\caption{\textbf{Fractional activation for varied threshold values.}  (\textbf{a-b})  The fractional activation calculated using a threshold value of 0.2 shown as a function of Functional Effect (a) and Structural Effect (b). (\textbf{c-d}) The same as (a-b) but using a threshold value of 0.4.  These results are comparable to that shown in Fig. 6c-d of the main manuscript.
\label{figS3}}
\end{figure}

\references{WC_bib}

%